\documentclass[sigconf,nonacm]{acmart}
\AtBeginDocument{%
  }

\usepackage{xspace}
\usepackage[capitalise]{cleveref}
\usepackage{tabularx}
\usepackage{booktabs}

\newcommand{\etal}{{\it et~al.}\xspace}
\newcommand{\gemini}{~Gemini\xspace}
\newcommand{\tripadvisor}{~Tripadvisor\xspace}

\usepackage{xcolor}

\usepackage{epigraph}
\usepackage{wrapfig}
\usepackage{graphicx}

\usepackage[most]{tcolorbox}

    \usepackage{multirow}

\begin{document}
\title{How Constraints and Preferences Shape Travel Planning: Implications for AI Planning Support}

\author{Fuling Sun}
\affiliation{
  \institution{University of California San Diego}
  \city{La Jolla}
  \state{California}
  \country{USA}
}
\email{fusun@ucsd.edu}

\author{Yining Cao}
\affiliation{
  \institution{University of California San Diego}
  \city{La Jolla}
  \state{California}
  \country{USA}
}
\email{yic069@ucsd.edu}

\author{Peiling Jiang}
\affiliation{
  \institution{University of California San Diego}
  \city{La Jolla}
  \state{California}
  \country{USA}
}
\email{peiling@ucsd.edu}

\author{Mingyi Li}
\affiliation{
  \institution{Northeastern University}
  \city{Boston}
  \state{Massachusetts}
  \country{USA}
}
\email{li.mingyi2@northeastern.edu}

\author{Haijun Xia}
\affiliation{
  \institution{University of California San Diego}
  \city{La Jolla}
  \state{California}
  \country{USA}
}
\email{haijunxia@ucsd.edu}

\begin{abstract}
Planning is a common yet complex activity shaped by constraints to satisfy and preferences to balance. Travel planning, as both an everyday activity and a frequent benchmark for evaluating intelligent systems, offers a rich context for examining how constraints and preferences emerge and evolve. While recent AI systems have achieved impressive results in generating personalized itineraries, they often assume that users can articulate stable goals upfront. To understand how real-world planning unfolds, we conducted a two-part interview study: one with eight travelers reflecting on their planning experiences, and one with nine travel agents sharing professional practices. We trace the dynamics of constraints and preferences as they are surfaced, refined, and coordinated throughout the planning process, and identified 11 actions revolving around constraints and preferences, which shaped the planning process. We offer design heuristics for planning tools that better support human–AI collaborative actions to support the fluid, contingent nature of planning.

% it often assumes users' inputs such as goals, requirements, .. are clear and ready to process 
% it often overlooks how users' inputs are formed. In human-centered tasks, constraints and preferences are not fully known at the outset, nor easily articulated. We focus on travel planning, a domain that involves open-ended goals, contingencies, and dynamic needs, and therefore representative for..... To investigate how people reason about planning, we conducted a two-part study: interviews with eight travelers reflecting on their planning experiences, and nine professional travel agents about their strategies. From this data, we trace how travelers construct their constraints and preferences through external, internal and structural factors, and three core dynamics on how users surface, coordinate and respond to them. Our findings echo prior work on the constructive nature of user needs and offer a detailed account of planning in practice. We discuss implications for planning tools that support articulation, adjustment, and flexible commitments. 

\end{abstract}




\maketitle

\section{Introduction}
% strating from planning and constraints 

% how previous algo use constraints to solve planning problem (classic planning) and later the ..
% LLM help those hard to formulated ones

% however, the assumption of knowing the constraints ...

% travel planning, is an example for planning that might not have clear input upfront....

% to take a detail look of how constraints and preferences ... 
% rq1: how constraints and prefernces shape travel planning

% result:
%  what are the constraints and preferences from 
%  the dynamics (surfacing and coordinating) 
%  the actions taken on the constraints and preferences  (moves)

% design heuristics / implications 
%   the model : constraints as ...
%  considerations?
%   the need: materialzie constraints 

%   the roles: coorindating, not satisfying

% how to jump in travel planning earlier?
% planning's importance and complexity?
% travel planning for example, is a everyday task from short term transportation to vacation arrangemtn... researchers studied the various aspect of it (list some from the tourism domain, decision making, ..

% researchers have studied how this process work and try to automate it. classic planning.... in factory... where spec are ready and clear.. recent AI have enabled to the more situated environment for robots, .... and for scenarios where 

% yet planning like travel planning, they are situated in ... contingency ... therefore, the form of the constraints and prefernces is dynamic and ... 

% therefore to study how constraint and prefernces shape travel planning to understand how supporting tools should be designed, we ....
Planning is a fundamental human activity that underlies how people organize actions, allocate resources, and negotiate trade-offs in everyday life and professional tasks~\cite{suchman1987plans, hayes1978cognitive}. Prior work in cognitive science and organizational studies has shown that planning is not purely a linear process of optimizing predefined goals, but an iterative activity shaped by knowledge, constraints, and situated judgements, aiming for plans that are both feasible and satisfying~\cite{simon1956rational, suchman1987plans, hayes1978cognitive}. 

Travel planning offers a particularly rich and relatable setting for examining these processes. From short-term transportation to extended vacation arrangements, travel planning is \textit{``temporal, dynamic, successive, multistage, and contingent''}~\cite{fesenmaier1990theoretical, jeng2002conceptualizing}. Travelers must coordinate logistical constraints such as time, budget, and availability, while also considering less tangible preferences such as comfort, interests, and desired experiences. These characteristics make travel planning both an everyday activity and a demanding coordination task, and a common showcase for intelligent planning and recommendation systems~\cite{google2025travelai, nyce2024}.

To support planning, classical automated planning supports rely on symbolic representations and goal specifications, enabling systems to generate action sequences that satisfy predefined conditions through optimization, constraint reasoning, or data-driven methods~\cite{fikes1971strips, ghallab2004automated, kaelbling1998planning}. Recent AI systems increasingly use large language models (LLMs) to generate plans from natural language descriptions, allowing users to express goals and constraints in flexible and conversational ways~\cite{cao2025large, pallaganiProspectsIncorporatingLarge2024, liHumanCenteredPlanning2023}. While these systems can produce seemingly plausible plans, they often assume that users can articulate stable goals and preferences upfront. This assumption simplifies system design but risks misaligning with how planning unfolds in practice and contrasts with established accounts of planning and decision making as an emergent and iterative process~\cite{bettman1998constructive, Smith_2021, suchman1987plans, deng2025retail}. 

This gap raises questions about how emerging AI-based planning tools should better accommodate the dynamic and iterative nature of planning for real-world tasks. Inspired by Suchman's work that argued that human plans and actions are outcomes shaped moment-by-moment by the environment \cite{suchman1987plans}, in this work, we focus on studying the shaping forces in the planning processes, using travel planning as a case study domain. Specifically, we treat \textit{constraints} and \textit{preferences} as the first-class objects in our investigation, and aim to understand:

% This gap raises questions about how emerging AI-based planning tools should better accommodate the formation and coordination of constraints and preferences in real-world tasks. In this work, we focus on traveling planning as a case study domain, and treat constraints and preferences as the first-class citizens in our investigation. By tracking them through the travel planning processes, we hope to understand

\begin{itemize}
    \item \textbf{RQ1}: How do constraints and preferences emerge and shape the travel planning process? 
    \item \textbf{RQ2}: What do the findings imply for the design of future AI-powered planning systems?
\end{itemize}

To gain understandings from different types of planning practices, we conducted a two-part interview study. First, we interviewed eight travelers about their recent travel experiences and planning processes. Then, we asked them to experience AI-powered tools to plan upcoming trips and share their feedback on incorporating AI into their travel planning. In the second part of the study, we interviewed travel agents to understand their strategies for assisting clients in travel planning tasks. 

From our interviews, we summarized the identified dynamics revolving around constraints and preferences, focusing on how they emerge and evolve over time (\textbf{RQ1}), such as how constraints and preferences are surfaced with information exploration, and how they are coordinated during planning to balance different tradeoffs. We distill 11 actions revolving around constraints and preferences, which reflect what travelers need, how travel agents help, and where current AI-assisted planning systems fall short. 

For example, we found that AI tools often rushed to generate itineraries, leaving travelers with generic plans, while travel agents employed a variety of strategies to elicit and probe travelers' preferences and arrive at more meaningful outcomes. Similarly, when consolidating diverse constraints and preferences into a final plan, AI systems rarely engaged in clarifying or resolving conflicts, instead, responding only to explicitly stated needs. In contrast, human travel agents actively identified ambiguities and facilitated negotiation to guarantee their understanding of travelers and compose feasible and personalized plans.

Moving forward from these empirical findings, we outlined design implications for planning tools to address the tensions between real-world planning dynamics and the assumptions embedded in current AI-assisted systems (\textbf{RQ2}). First, we highlighted how system design should model constraints and preferences as evolving representations that remain visible and manipulable throughout planning, allowing users to revisit earlier inputs, adjust priorities, and explore alternative plan configurations as their understanding develops. Second, for each of the 11 identified actions, we suggest human-AI collaborative actions on constraints and preferences that allow users and agents to articulate, refine, and reconcile them over the course of planning. 
% These collaborative actions, including specifying, refining, and detecting conflicts, help establish a shared understanding between users and systems of which constraints and preferences are active and how they should shape the plan. 

In summary, the contributions of this work include:
\begin{itemize}
    \item \textbf{An empirical investigation} of how constraints and preferences are surfaced, coordinated, and negotiated in the context of travel planning, based on interviews with travelers and professional travel agents.We identify 3 categories of constraints and preferences, distill 11 recurring actions centered on them, and reveal key tensions between real-world planning practices and the assumptions embedded in current AI-assisted planning tools.
    \item \textbf{Design implications} for AI planning tools, detailing how constraints and preferences should be modeled as evolving representations and how systems should support \textit{human-AI collaborative actions} for surfacing, refining, and reconciling them throughout planning.
\end{itemize}

\section{Related Work}

\subsection{Understanding Planning and Travel Planning}
\label{rw:ptp}
%   more detailed: why dynamics, why non-linear
%   classic: linear, step by step, ...

% understanding planning from the psychology perspectives
% 
Planning is a fundamental cognitive activity humans use everyday to generate courses of actions to achieve goals~\cite{hayes1980human,hoc1988cognitive,korf1987planning}. Early research in cognitive science and psychology analyzed planning from various perspectives, such as problem-solving contexts (e.g, Tower of Hanoi) and everyday tasks (e.g, errand planning), to understand how people generate sub-goals, adjust their strategies~\cite{hayes1978cognitive, anzai1979theory, miller1960plans,simon1971human}. This line of work led to influential models such as the Test-Operate-Test-Exit (TOTE) unit, a feedback loop of evaluation and adjustments of actions~\cite{miller1960plans}. Studies of everyday planning further demonstrated its \textit{``opportunistic, incremental, multidirectional, and heterarchical''} nature~\cite{hayes1978cognitive}. Collectively, these studies emphasized that planning is a dynamic process of constructing, evaluating, and revising actions to achieve goals.
% , supported by mental representations of the problem space~\cite{simon1971human, hoPeopleConstructSimplified2022}

To support planning, researchers have investigated and proposed various approaches in different domains. One line of work considers planning as the generation of scripts of actions toward a known end~\cite{fikes1971strips,russell2016artificial}. Within this framework, planning problems are represented formally and solution are computed through search or optimization. Domains such as manufacturing, business process management, and logistics have applied automated planning to support decision making and to increase efficiency, given the relatively clear specification of inputs and outputs~\cite{bourne2011recent, marrella2019automated, karpas2020automated}. In this type of planning, constraints are assumed to be explicitly known, or at least representable within the predefined model, and dynamics are handled as updates to the model rather than as emergent conditions.

Another line of work emphasizes planning as situated action. Suchman argued real-world actions often deviate from pre-designed plans and must be adapted \textit{``with respect to specific, local, contingent determinants of significance''}~\cite{suchman1987plans}. From this perspective, \textit{``plans are the resources for situated actions''} rather than prescriptive instructions, orienting actors' behaviors but not fully determining their conduct. Domains that involve decision making under uncertainty (e.g., clinical decision making in health care \cite{helou2020uncertainty}) and rapidly changing conditions (e.g., emergency response in improvisation \cite{mendonca2007cognitive}) illustrate how planning is shaped by emergent contingencies and situated in practice. 

Travel planning is another example of planning with situated actions. It involves interdependent decisions across destinations, transportation modes, accommodation, time frame, and activities~\cite{grigolon2013facet}, often under personal constraints and changing external conditions. Prior work describes it as a  \textit{``temporal, dynamic, successive, multistage, and contingent''} planning process~\cite{stewartCaseBasedApproachUnderstanding1999,fesenmaier1990theoretical, jeng2002conceptualizing}. Moreover, travel planning often involves multiple stakeholders, competing preferences, risks and uncertainties~\cite{johnston2012didn,garg2015travel,seabra2013heterogeneity}.
These characteristics make travel planning an ideal case for studying how constraints and preferences are surfaced, negotiated, and revised in context. 
Therefore, we use travel planning as a representative domain to investigate the dynamics of constraints and preferences, and to inform the design of systems that better support situated, open-ended planning.

% how do they study?

% formalize planning models for automation (why, how, ..)

% and why travel planning?

\subsection{Planning with Large Language Models}
% planning in various domains, so we can talk abou why ai travel planning can apply to others
Recent advances in artificial intelligence (AI), especially large language models (LLMs), have offered new opportunities to support planning tasks across diverse domains given their ability to interpret natural language and leverage large amounts of open-world knowledge~\cite{pallaganiProspectsIncorporatingLarge2024,ding2023integrating}.  Researchers have explored various approaches to leveraging LLMs for planning, including task decomposition, multi-plan selection, external planner for evaluation, reflection and refinement, and memory aid~\cite{huang2024understanding}, alongside prompting strategies such as few-shot learning~\cite{song2023llm} and chain-of-thoughts~\cite{wei2022chain,stechly2024chain}. 
To evaluate whether LLMs can accomplish complex planning tasks in realistic settings, researchers have evaluated LLMs in various planning domains, including the blocks world problem, travel planning, meeting planning, and calendar scheduling~\cite{stechly2024chain, xieTravelPlannerBenchmark2024, zheng2024natural,deng2025retail}. However, performance remains limited, in part due to the narrow or implicit treatment of constraints and the diminishing influence of user questions or preferences as planning unfolds~\cite{xie2024revealing}.

Alongside improving LLMs' performance to achieve human-level intelligence, researchers have also been exploring how to integrate human feedback with AI systems, and investigate design patterns that support a smooth, collaborative experience in domains requiring planning and reasoning, such as writing, data analysis, and travel planning~\cite{reza2025co, kazemitabaar2024improving, ma2024beyond}.
Engaging users during the process often leads to better outcomes. For example, when the inquiry is under-specified or no feasible solution is available, systems can use proactive clarification to elicit more information~\cite{zhang2024ask}. To support user verification and editing, recent systems present AI-generated plans in visual representations. For example, Kazemitabaar\etal introduced generative data analysis tools that allow users to inspect, edit, and steer generated outputs~\cite{kazemitabaar2024improving}. Similarly, WaitGPT visualizes plans in node-link diagrams to help users understand their structure, revise steps more easily~\cite{xie2024waitgpt}.

For more exploratory tasks, researchers focus on the collaboration and dynamics between humans and AI to support open-ended planning and flexible decision-making~\cite{kim2025plantogether, feng2024cocoa,reza2025co}. For example, COCOA enables users and AI to collaboratively compose and execute a research plan through an interactive interface~\cite{feng2024cocoa}. Other tools help users navigate large and complex problem spaces by offering guidance, or scaffolding the exploration~\cite{ma2024beyond, peng2025navigating}. For tasks that unfold over longer periods, users' needs often evolve and shift. To address this, Peng~\etal introduced a Need Panel that helps users organize their developing needs throughout the exploratory searching~\cite{peng2025navigating}. While recent systems allow users to add, remove, or edit needs~\cite{peng2025navigating}, or adjust their flexibility and priority~\cite{lee2025veriplan}, the coordination of multiple, potentially conflicting needs remains under-explored. 

Yet in real-world tasks like travel planning, managing the interplay among diverse constraints and preferences is inevitable. As AI-assisted travel planning is increasingly used in practice, researchers have begun examining how people interact with and adopt these tools, including factors such as trust, ease of use, and expectation gaps in AI-generated travel recommendations~\cite{ali2023antecedents, topsakal2025familiarity,sigala2024understanding}.  We build on this line of work by focusing on how constraints and preferences are handled in context, informing the design of human–AI collaborative planning systems.
Therefore, we rooted from travel planning scenarios and interviewed travelers and travel agents' practices, to learn their practices in handling the complex and evolving needs to inform the design of human-AI collaborative planning systems.

\subsection{Constraints, Preferences and Personalization}

% hmm where to put recommender system

% personalization, e.g. recommender systems..
%   some technique they use to capture users preferences ...

% the dynamic of constraints..
%   so.. elicit users' unknown prefrences..
%  

Personalization is a central goal for many AI-assisted systems, aiming for results that align with users' needs and goals, rather than one-size-fits-all outputs~\cite{knijnenburg2012explaining}.
Research shows that users' constraints and preferences are rarely fixed, but follow a constructive nature~\cite{bettman1998constructive}. They can be under-specified at the outset, emerge during the process, shift in priority over time, and sometimes cause conflicts with other needs~\cite{payne1993adaptive, sanna2021next}. 

Therefore, to meet individual users' evolving needs, systems have to go beyond capturing the static profile of users. Prior research has explored methods for inferring or eliciting users' preferences. Recommender systems infer preferences from behavioral traces such as clicks, ratings, often using collaborative filtering~\cite{su2009survey} or content-based methods~\cite{Meteren2000UsingCF} to predict what a user may like~\cite{sanna2021next, chen2004survey}. On the interaction side, prior work has developed approaches that support users articulating and refining their preferences interactively, such as structured exploration~\cite{ma2024beyond} and example-critiquing~\cite{viappiani2006preference}. 

% Recent work on human–AI alignment similarly relies on user preferences over LLM outputs~\cite{stiennon2020learning}, and has inspired interface designs with visualizations to support quicker and more reliable judgments~\cite{shi2025dxhf}.

Users' preferences and constraints, ranging from hard and soft constraints to qualitative conditional preferences, play distinct roles in personalization~\cite{Domshlak2006-cq}. Moreover, these inputs are often dynamic, shifting based on evolving user states or contextual factors~\cite{lyu2021workflow, tanca2011problems}. Beyond their individual roles, constraints and preferences often interact, where satisfying one condition reshapes the relevance or priority of others. Research in constraint satisfaction and multi-criteria decision making has long emphasized the need to reason about interdependencies among constraints, seeking optimal trade-offs that balance competing objectives~\cite{domshlak2006hard, keeney1993decisions}. 

While these formal models aim at identifying an optimal solution, many everyday activities such as personal scheduling or travel planning have no single correct answer~\cite{hayes1978cognitive,simon1956rational}. Rather than optimizing, people often adaptively balance constraints and preferences, negotiating trade-offs to reach outcomes that feel feasible and satisfying~\cite{simon1956rational}. Interface designs can make these negotiations more visible, helping users inspect, compare, and reason across multiple dimensions. For example, Bilbily~\etal integrated a map, a calendar, and lists in a mobile application to support users in arranging tasks and resolving temporal and spatial conflicts with linked visualizations~\cite{bilbilySpaceTimeChoice2021}. Similarly, Embark supports travel planning by consolidating structured data (e.g., dates), rich views (e.g., maps), and spreadsheet-like formulas into a single interactive document, reducing the need to switch between multiple applications~\cite{sonnentag2023embark}.  

We aim to build on these findings by focusing on the dynamics and flexibility of how constraints and preferences engage in planning. Our goal is to understand how constraints and preferences are formed, and how they guide planning tasks, and to explore how interface design, in combination with AI, can make these dynamics more visible and actionable.

\section{Methods}
% We chose travel planning as our focus domain because it exemplifies open-ended, real-world planning shaped by evolving needs, as discussed in Section \ref{rw:ptp}. 
To answer the research questions in the context of travel planning, we decided to seek insights from two perspectives: experienced travelers who plan their own travels and professional travel agents who plan travels for their clients. Studying both groups allowed us to capture both end-user planning experiences and expert strategies for assisting others in travel planning. 

% First, we conducted an interview study with frequent travelers. By learning from travelers, we aimed to identify common practices and understand the rationale behind these behaviors. Additionally, we asked participants to try using AI systems to plan a trip. From their experiences, we identified the strengths and weaknesses of current AI system designs. Next, we interviewed experienced travel agents to uncover their strategies for assisting with travel planning. These strategies from domain experts offer valuable insights for improving AI system design. 

% From both perspectives, we gained insights into the types and dynamics of constraints and preferences in the planning process. By examining the practices of travelers and travel agents, we identified how these factors emerge and evolve, and categorized the pitfalls involved in managing them using current tools.

% Fro mthe two parts of study, we gained insights into three approaches to travel planning: planning by travelers, planning with AI systems, and planning with travel agents. These perspectives provide a comprehensive foundation for understanding how to design AI tools for planning tasks.

\subsection{Perspective 1: Learning from Travelers}
First, we conducted an interview study with frequent travelers. By learning from travelers, we aimed to identify common practices and understand the rationale behind these behaviors. 

\subsubsection{Participants}
We recruited eight participants (P1---P8; 4 male, 3 female, 1 prefer not to say; 21---38 years old) using convenience sampling through our institution's community channels. All participants had taken at least one trip in the past three months and had upcoming trips within the next three months at the time of the study (2024 summer). Participants reported various reasons for their travel, including visiting family or friends, taking weekend getaways, going on holiday, and attending conferences or business meetings. 

All participants had experience using AI tools in various domains, such as coding (e.g., GitHub Copilot), writing assistance (e.g., ChatGPT), and AI-enhanced searching (e.g., Google, Bing). 
As the concept of travel planning with AI tools was still in its early stages at the time of the study, none of the participants had utilized AI tools to plan their previous trips. Yet, it is worth noting that this study was not designed to evaluate expert use of AI tools or participants' prompting skills. We aimed to observe how participants engaged with these systems in ways that reflect their practices and provide feedback on this mode of planning. Therefore, participants' familiarity with the concept of AI served primarily to ensure baseline comfort with the technology, rather than to test their ability to use AI.

\subsubsection{AI System Selection}
To provide participants with a concrete experience for their feedback on AI-assisted travel planning, we selected two representative systems with distinct interactions styles. We selected \gemini (Gemini Pro) as a representative example of conversational interfaces for its versatile usage and integration with commonly-used applications (e.g., Google Maps), and \tripadvisor, representing a GUI-based domain-specific application.

\begin{figure*}
    \centering
    \includegraphics[width=\linewidth]{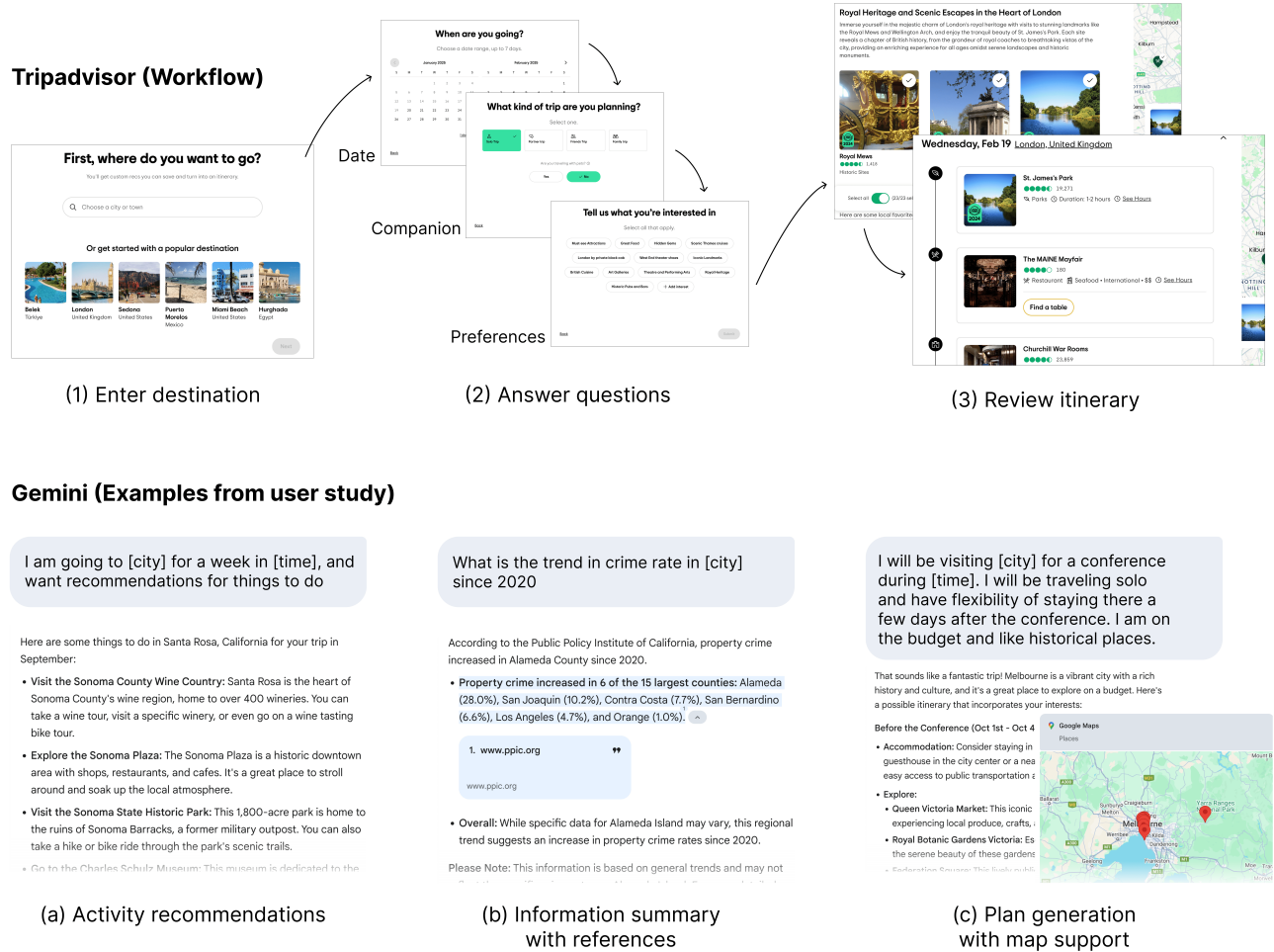}
    \caption{Demonstrations of \tripadvisor's workflow and \gemini's use cases. In \tripadvisor, users follow a structured process: (1) entering their destination, (2) answering questions about travel dates, companions, and preferences, and (3) receiving recommendations for points of interest, which are organized into a day-by-day itinerary. \gemini allows users to ask questions in various forms, such as (a) requesting category-specific recommendations, (b) seeking detailed information, and (c) asking for a complete plan.}
    \label{fig:tools}
    \Description{Figure 1: Composite figure comparing a structured travel-planning workflow in Tripadvisor (top row) with open-ended travel assistance in Gemini (bottom row). Top row shows three UI screenshots connected by arrows: (1) a destination entry screen with a search bar and popular destination tiles; (2) a sequence of question screens collecting trip date, travel companions, and interest/preferences; and (3) a generated itinerary view with a day header, a list of recommended places/activities, and an accompanying map. Bottom row shows three Gemini examples from a user study: (a) a prompt requesting things to do in a city and a response listing activity recommendations, (b) a prompt asking for a city crime-rate trend since 2020 and a response summarizing the trend with citations/references, and (c) a prompt requesting a budget, solo conference trip plan and a response that produces an itinerary with an embedded map and location markers.}
\end{figure*}

Our goal in including these tools was not to compare system performance, but to prompt participants to articulate and reflect on how their constraints and preferences were expressed and handled when interacting with AI planning tools. The usages of these two systems are demonstrated in~\cref{fig:tools}.

% One system was conversational interfaces, which included both general-purpose chatbots (e.g., \gemini) and travel-specific chatbots (e.g., MindTrip\footnote{https://mindtrip.ai/}). 
% When using chatbots, users interact with the tool by providing prompts in natural language. The other category of systems included graphical user interfaces (GUI) with AI-powered planning support for travel-related activities (e.g., \tripadvisor). With \tripadvisor, for example, users are guided through a series of travel-related questions (e.g., destination, duration, and preferences), after which they receive an AI-generated plan. 

\subsubsection{Procedure}
Each interview began after participants acknowledged consent and then proceeded in two phases. During the first phase, participants were asked to share their recent travel experiences and describe their trip-planning processes. They were also encouraged to share their itineraries and planning materials with the interviewer, as previously suggested in the sign-up form and confirmation emails. This phase lasted approximately 30 minutes. 

In the second phase, participants interacted with the two AI-assisted tools to gain hands-on experience with AI-assisted travel planning.
Rather than conducting full planning sessions, this phase was designed to give participants a concrete experience of using AI systems to plan trips and to prompt reflection on how these tools supported or constrained their planning practices. Participants selected destinations they had been considering traveling to and used one AI tool at a time to explore possible features and results. 
% While completing the tasks, participants were free to access any other tools as supplements. Participants first used one of the two AI tools for 30 minutes to plan an upcoming trip. 
They were asked to think aloud~\cite{lewis1982using}, and the interviewer occasionally asked questions if notable behaviors were observed. 
After trying the first AI tool, participants were asked to share their feedback regarding their planning experience and the result. Participants then used the second AI tool to plan a trip. After using this second AI tool, they were again asked to share feedback about their experiences. 

We note that the time allocated for each tool interaction (approximately 30 minutes) was shorter than real-world travel planning, which often unfolds over days or weeks. Nevertheless, participants reported that this duration was sufficient for them to develop a clear understanding of the distinctive planning characteristics of each tool. Participants were instructed to proceed at a comfortable pace, without the expectation of producing a complete itinerary. Accordingly, the goal of this phase was not to evaluate the quality of the resulting plans, but to examine how participants articulated, negotiated, and revised constraints and preferences while engaging with AI-assisted planning tools. To mitigate potential order effects, the presentation order of the AI tools was randomized across participants.

% As the time allocated to use each AI tool was shorter than the duration typically spent on real-life travel planning, which often spans weeks and even months, participants were instructed to proceed at a pace that felt comfortable to them without needing to produce a complete travel plan within the allotted time. 

% This time, albeit short, is sufficient for our study purposes, as our goal is not to ask participants to produce a final plan, but to gain an experience of using AI tools, and see whether these AI tools handle participants' constraints and preferences. The order of AI tools in studies was randomized across participants to mitigate potential order effects.

All interviews were conducted and recorded through Zoom. Participants were asked to share their screens with the interviewer to facilitate the discussion. Each interview lasted approximately 90 minutes. As compensation, participants received a \$40 Amazon gift card after the interview.

\subsection{Perspective 2: Learning from Travel Agents}
Next, we interviewed experienced travel agents to uncover their strategies for assisting with travel planning. 
This allowed us to study how constraints and preferences were addressed in expert-mediated planning.
% These strategies from domain experts offer valuable insights for improving system design. 
\subsubsection{Participants}
To recruit domain experts, we curated a list of travel agents from an industry catalog and our institution's partner travel agencies. We sent invitation emails to the travel agents with positive online reviews, stating their active professional practice and strong client feedback. Nine travel agents agreed to participate in the interviews (T1---T9; 8 female, 1 male; ~\cref{table:advisors}). The gender imbalance reflected the distribution of travel agents, a profession that is predominantly female~\cite{datausa_travelagents}.
Among them, two specialized in corporate travel (i.e., trips taken by employees for business purposes), while the remaining focused on leisure travel (e.g., tours, cruises, and honeymoons). 

\begin{table}[]
\caption{Demographic information about the interviewed travel agents, including their self-reported travel specialty. }
\label{table:advisors}
\begin{tabular}{lcccc}
\hline
ID & Age Range & Gender & Work Experience& Specialty                                         \\ \hline
T1  & 75-84     & Female & 23 years           & Leisure\\
T2  & 55-64     & Male   & 34 years           & Leisure\\
T3  & 35-44     & Female & 2 years            & Leisure\\
T4  & 55-64     & Female & 20 years           & Leisure\\
T5  & 55-64     & Female & 40+ years          & Corporate\\
T6  & 35-44     & Female & 20 years           & Leisure\\
T7  & 45-54     & Female & 11 years           & Leisure\\
T8  & 45-54     & Female & 28 years           & Corporate                                                                  \\
T9  & 35-44     & Female & 7 years            & Leisure\\\hline                                                              
\end{tabular}
\Description{The table presents demographic information about nine interviewed travel agents, detailing their age ranges, genders, work experiences, and self-reported travel specialties. The advisors' ages vary, with one in the 75-84 range, three in the 55-64 range, two in the 45-54 range, and three in the 35-44 range. The group comprises seven females and two males, with work experience ranging from 2 years to over 40 years. Most of the advisors, specifically seven out of nine, specialize in leisure travel, while the remaining two focus on corporate travel.}
\end{table}

\subsubsection{Procedure}
We conducted semi-structured interviews, with two held in-person and seven held remotely via Zoom. After briefly introducing the interviewer and the research project, we began by asking the travel agents to share a recent plan they created for a client. Based on this example, we explored their processes of working with clients, including how they communicated with their clients, how they created customized itineraries, and their reflections on their role as travel agents. 
After answering each question, we asked the travel agents if similar situations had occurred beyond the initial example to gain a broader understanding of their practices through various examples. Additionally, the travel agents were asked to share, if possible, relevant documents, itineraries, or other materials they used, and explain how they created and shared them with clients. Each interview lasted 60---90 minutes, and the travel agents received an \$80 Amazon gift card as compensation. 

% \subsubsection{Analysis}

% All interview recordings were transcribed and analyzed using Condens. We employed both deductive and inductive thematic analysis~\cite{braun2019reflecting} to identify key themes from the interviews. After familiarizing themselves with the interviews, the first author coded two interviews to develop preliminary codes, which were then discussed and refined with the other authors. In the second coding round, the first and fourth authors collaboratively coded the interviews using the established code list, revising and refining the codes as new insights emerged through discussion. Once all nine interviews were coded, the first author conducted a final review of all interviews to identify any missing or misinterpreted data and made necessary updates to the coding. Finally, the first author analyzed all excerpts and grouped the codes to generate overarching themes. 
% Following further discussion and refinement, three major theme groups with eight themes were identified, highlighting their insights into working as agents and their best practices.

\subsection{Analysis}
All interviews were transcribed and analyzed using Condens\footnote{https://app.condens.io/}. We first analyzed traveler and travel agent interviews separately, using iterative coding to surface patterns in how constraints and preferences were expressed and negotiated. Codes were treated as provisional and were revised as interpretations developed through analytic discussions among the research team. The first author organized codes into candidate themes, which were reviewed and refined through team discussions to capture coherent patterns within each participant group. We then synthesized themes across both participant groups, comparing patterns and refining overarching themes to capture shared dynamics and contrasting perspectives in travel planning.
% We conducted the initial thematic analyses separately on the two sets of interview data, allowing themes to emerge inductively within each perspective. 
% Then, to create a cohesive understanding, we synthesized the identified themes from the two studies within a unified context, linking and categorizing findings into overarching themes and subcategories to facilitate comparison and highlight shared insights and contrasts. 

\subsection{Limitations}
% \todo{still deciding if this should be here}
The study design has several limitations. First, we used convenience sampling, and our participants may not represent the full diversity of travel planning practices, particularly across different cultures, socioeconomic backgrounds, and travel styles. Second, although participants interacted with AI planning tools during the interviews, these interactions were brief and did not fully capture the extended and iterative nature of real-world travel planning, which often unfolds over days or weeks. As a result, our findings reflect snapshots of planning behavior and reflection rather than complete end-to-end planning processes.
Despite these limitations, our study still provided an empirical basis for examining constraints and preferences in travel planning and informing the design of AI-supported planning tools.

% \section{Results}
\section{Findings: The Dynamics of Constraints and Preferences in Travel Planning}
% Both our interviews and prior work~\cite{lyu2021workflow, tanca2011problems} reflect that constraints and preferences are rarely fixed, neither at the outset nor throughout the process. 

% Instead, they shift as travelers gather and compare options, negotiate trade-offs, and make commitments. 

% To better understand the changes, along with the practices and challenges they entail, we organize our findings into two phases: \textbf{surfacing} and \textbf{coordinating}, which capture how travelers encounter and work with constraints and preferences as planning unfolds.
Our findings highlight the dynamic nature of constraints and preferences, summarized through 3 types of sources from which they arise, and 11 recurring actions through which they are surfaced, applied, and influence planning.

\subsection{Types of Constraints and Preferences}
Travel planning involves a series of decisions. Learning from travelers and travel agents, we categorize the constraints and preferences used to make decisions into three types, depending on their sources.

\subsubsection{External Factors: Destination Constraints and Offerings}
\label{external-factors}
% external factors (e.g., local conditions, cultural norms, seasonal offerings) 

In travel planning, the plan will ultimately be executed by travelers within a specific time and place. Therefore, what is available in the destination, such as accommodations, attractions, and events, as well as modes and schedules of transportation, will directly shape what is feasible. These \textit{external factors} originate from the environment in which the plan is enacted, usually are constraints that exist independently of the travelers. Sometimes, significant decisions are made due to external factors. For example, P4 changed her travel dates to catch the whale-watching season.

Without considering \textit{external factors}, travelers might miss constraints or have an unrealistic mindset for the eventual results. Travel agents mentioned that even though they strive to fulfill clients' wishes for their trips, occasionally, clients may propose whimsical or impractical ideas. For example, during
the 2024 Olympics, T3 had a group of clients that \textit{``just bought their ticket to Paris but not wanting to go to the Olympics, just sightseeing. But you can't. You're not gonna be able to go to the Eiffel Tower when Céline Dion is singing there.''} Similarly, T6 shared her experience with clients underestimating travel logistics:

\begin{quote}
 \textit{``clients will say, `I want to go to the Maldives, but I only have four days.' [They are] not able to do it because just the flight to get you to the Maldives is going to take you [days]. So sometimes people don't understand travel distances.''}  (T6)
\end{quote}

\subsubsection{Internal Factors: Travelers' Constraints, Preferences, and Intentions}
Travel is a significant investment of time, effort, and money. Consequently, travelers expect their plans to reflect their individual likes and dislikes. These \textit{internal factors}—personal constraints, preferences, and goals—not only shape the construction of a plan and its feasibility, but also determine whether the resulting experience is meaningful and memorable from the traveler’s perspective.

% why it is important
Various specific factors stem from the travelers, such as who they are traveling with, the purpose of the trip, and their past experiences. However, even when these factors appear similar on the surface, they can shape planning in very different ways across individuals. For example, although both P7 and P8 were traveling to a city they had visited before, their intentions diverged: P7 expressed a desire to revisit places that \textit{``make me reminisce [about] the good old days when I visited there''} whereas P8 preferred to avoid areas she had already explored. These differing attitudes resulted in contrasting reactions when familiar locations were recommended by the AI tools.

\subsubsection{Integrative Factors: Coherence Across the Plan}
\label{integrative-factors}
As plans take shape, individual components introduce another layer of consideration: the structure of the plan itself. \textit{Integrative factors} describe how activities are interconnected. They create anchors that add to or reshape existing options, often constraining what comes next. For example, booking a flight or hotel establishes fixed temporal and spatial anchors, which in turn limit how other activities can be arranged. The interdependency determines how well the overall plan holds together as a coherent and executable whole.

Without attention to such integration, travel plans can become unbalanced. For instance, when reviewing an AI-generated itinerary, P1 criticized the inclusion of four museum visits across just three days, despite only expressing a general interest in museums.  Travelers also emphasize the importance of pacing and variety across activities. As P4 explained, she designed her trips by balancing \textit{``high energy activities in the morning, low energy activities in the evening, things like that.''} These examples highlight how \textit{integrative factors} involve not only structural dependencies but also judgments about rhythm, balance, and coherence across the plan.

\subsubsection{Interplay among Factors}
These three types of factors identified from the interviews are not meant to be mutually exclusive, exhaustive, or fixed. Instead, we use them to understand how different considerations arise, evolve, and guide planning. A single factor may span or shift across categories based on the context. For example, when P1 mentioned his goal to watch the sunset in the park, it began as an internal intention, became tied to external conditions such as sunset timing, and eventually imposed integrative constraints on how other activities were arranged. Importantly, such interplay does not always follow a clear or linear transition. Our aim in proposing this categorization is not to rigidly classify each factor, but to surface how different types of constraints and preferences interact throughout the evolving planning process, which leads to the following findings. 

\subsection{Actions Revolving around Constraints and Preferences}
\label{set:actions}

\begin{table*}[htbp]
  \centering
\caption{Actions organized into three groups: surfacing constraints and preferences, applying constraints and preferences to plans, and managing decision space with constraints and preferences. For each action, the table summarizes its role in planning, who currently performs it, and gaps in current AI planning support.}
  \label{tab:action}
  \includegraphics[width=\textwidth]{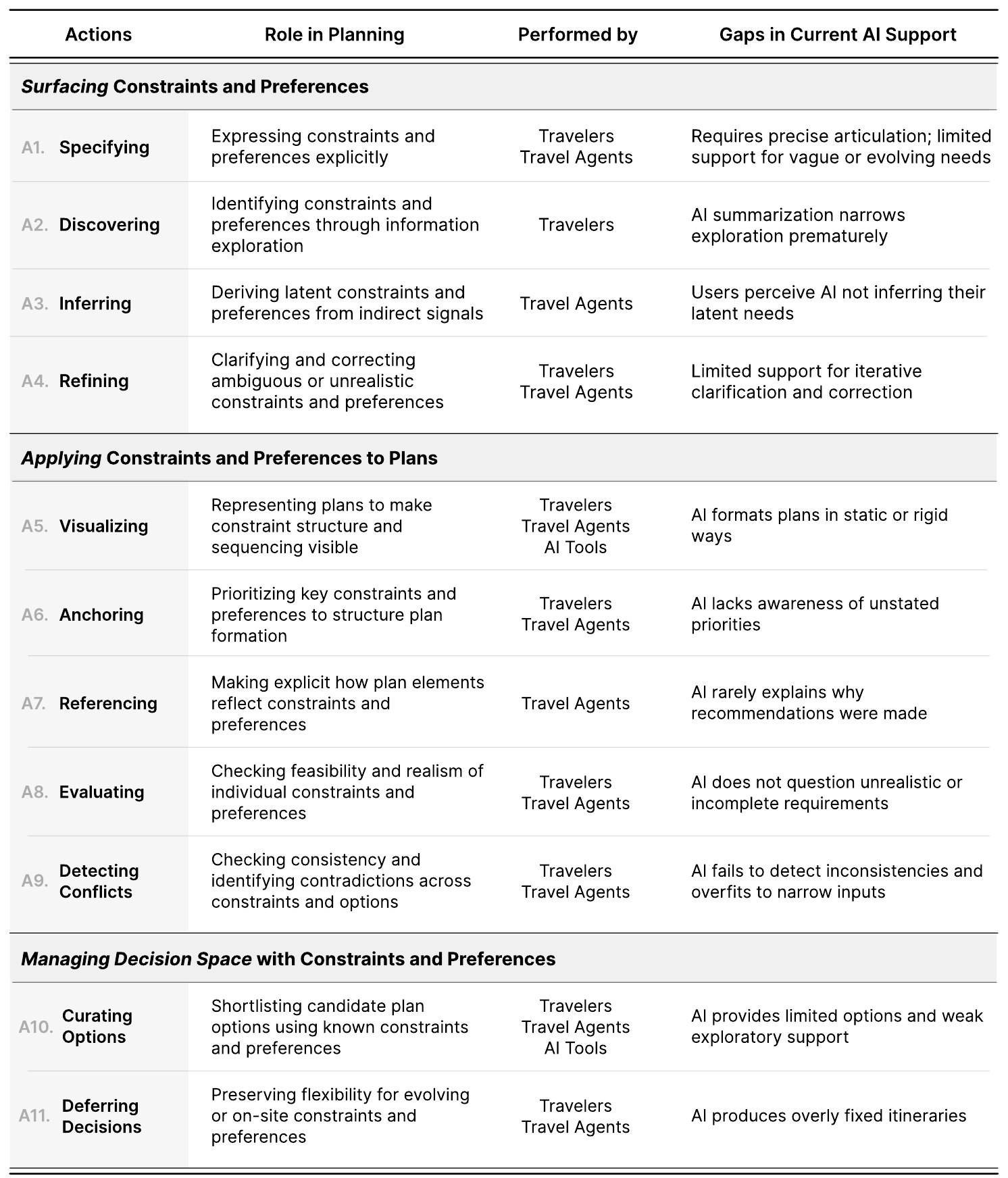}
\Description{Table 2: Table 2 summarizes 11 planning actions (A1--A11) organized into three groups, with four columns: Actions, Role in Planning, Performed by, and Gaps in Current AI Support. Surfacing Constraints and Preferences includes A1 Specifying (explicitly expressing constraints/preferences; performed by travelers and travel agents; requires precise articulation and has limited support for vague/evolving needs), A2 Discovering (finding constraints/preferences through exploration; travelers; AI summarization can narrow exploration prematurely), A3 Inferring (deriving latent needs from indirect signals; travel agents; AI does not infer latent needs), and A4 Refining (clarifying/correcting ambiguous or unrealistic constraints; travelers and travel agents; limited iterative clarification/correction support). Applying Constraints and Preferences to Plans includes A5 Visualizing (representing plans so constraint structure and sequencing are visible; travelers, travel agents, AI tools; AI formats plans in static/rigid ways), A6 Anchoring (prioritizing key constraints to structure plan formation; travelers and travel agents; AI lacks awareness of unstated priorities), A7 Referencing (making explicit how plan elements reflect constraints/preferences; travel agents; AI rarely explains why recommendations were made), A8 Evaluating (checking feasibility/realism of individual constraints/preferences; travelers and travel agents; AI does not question unrealistic or incomplete requirements), and A9 Detecting Conflicts (checking consistency and contradictions across constraints/options; travelers and travel agents; AI fails to detect inconsistencies and overfits to narrow inputs). Managing Decision Space with Constraints and Preferences includes A10 Curating Options (shortlisting candidate plan options using known constraints/preferences; travelers, travel agents, AI tools; AI offers limited options and weak exploratory support) and A11 Deferring Decisions (preserving flexibility for evolving or on-site constraints/preferences; travelers and travel agents; AI produces overly fixed itineraries).}
\end{table*}

To understand how constraints and preferences shape the planning process (\textbf{RQ1}), we summarized our findings based on the actions performed on or related to constraints and preferences during the planning process. For each action, we reported how it was carried out in current practices, whether by travelers, travel agents, or AI systems, and highlighted the challenges or gaps when the action was poorly supported. \cref{tab:action} summarizes these actions. 

This set of actions is not intended as a comprehensive taxonomy of all possible operations on constraints and preferences, due to the limited scope of our study subjects. Rather, it serves as an analytic lens grounded in our empirical observations, highlighting what kinds of work are required around constraints and preferences, what strategies humans employ to perform planning, and where current AI systems fall short.

\subsubsection{\textbf{A1: Specifying Constraints and Preferences}}
 This action captures moments when travelers explicitly state constraints and preferences. Direct specification requires travelers to translate their intentions into precise expressions that others or systems can interpret when co-planning. This operation turns subjective needs into concrete parameters that can be immediately applied to filter or generate plan options.

In the interviews, travelers performed direct specification when using filters on travel websites, onboarding questionnaires, or free-text queries in search engines and AI chatbots. Travel agents similarly attempted to collect direct specifications through intake forms at the beginning of client collaborations. However, agents reported that these forms were often incomplete or ignored. As T3 noted, \textit{``people didn't really like filling out forms''}, yet travel agents still sent them out to let \textit{``they know what to expect''} (T6) for the later conversations. 

Because travel planning is inherently exploratory, participants often struggled to formulate precise requests when their intentions were still unclear. As P3 described, \textit{``sometimes I don't know what I am trying to ask''}. 
Many \textit{internal factors} related to personal likes, dislikes, and 
habits, were nuanced and hard to express directly. As P8 noted, even when she had a sense of the kind of hotels she wanted, it was hard to translate into keywords or filters, \textit{``I think it is a rather abstract concept, so I don't know how to tell the system what I want.''} Rather than specifying these preferences to AI systems, she preferred to browse detailed options and make her decisions, which allowed her to refine and recognize what matched her needs.

\subsubsection{\textbf{A2: Discovering Constraints and Preferences through Exploration and Examples}}
In this action, travelers discover constraints and preferences by inspecting concrete examples. During planning, travelers discover what is possible and what they might want through exposure to detailed itineraries, shared experiences, and peer-generated content, grounding abstract intentions in detailed examples.

From the interviews, we learned that travelers performed this action by drawing on a wide range of online resources, from general introduction to a city to details about specific sights and dining options. 
% This intake of information not only shaped travelers' understanding of the location (\textit{external factors}), but also inspired their personal preferences (\textit{internal factors}), especially for unfamiliar locations and foreign destinations. 
One common strategy reported by travelers was to look for travel plans shared by people with a similar background. Such shared content provided both an overview of a practical itinerary with detailed arrangements (\textit{integrative factors}) and guidance on the considerations to take into account from a perspective closer to their own, which surfaced a wider range of factors:
\begin{quote}
\textit{``I tend to look at people who are similar to me, like I'm going traveling with my parents so then look for people traveling with parents. [They] tend to do these things on these days. I can trust them a bit more because they tend to also have individual opinions, similarly aligned preferences and constraints probably.''} (P4)
\end{quote}

Travel agents performed a similar action by reusing previously constructed itineraries to help clients quickly learn what was feasible at a destination. Concrete examples enabled clients to accumulate understanding of available options and to recognize their own likes, dislikes, and practical constraints through comparison with detailed plans.

Compared to browsing the Internet extensively, the two AI systems provided travelers with a summary of the vast information space. For example, \gemini synthesized information from various resources, removing the need to manually collect information, as P6 noted:

\begin{quote}
    \textit{``If I do research by myself, I need to check each place and see how long it takes for other people to visit those places. It's a lot of work. But here I can just ask \gemini to give me a summarization.''} (P6)
\end{quote}

% the example of AI
However, participants noted that such summarization also constrained exploration. Compared to open-ended online browsing, AI systems provided only a small number of recommendations per request. Participants noted that this constrained exploration, especially in early planning stages when goals and preferences were still forming. As P3 reflected, jumping too quickly into personalized recommendations reduced his learning about the destination:

\begin{quote}
    \textit{``[Gemini] removed everything else I did not know, ... but on Google [Search], the zoo seems to be like a very famous thing ... I would have never known that there is a famous zoo [in the destination] using AI.''} (P3)
\end{quote}

\subsubsection{\textbf{A3: Inferring Constraints and Preferences through Past Experiences}}
In this action, constraints and preferences are surfaced by eliciting and interpreting past experiences. Rather than relying on travelers to specify requirements directly, it uses narratives of previous trips to reveal internal priorities, habitual patterns, and valued experiences that can guide future planning.

Travel agents performed this action by prompting clients to recount memorable moments from past travels, especially when clients were unfamiliar with the current destination. Instead of expecting precise preferences upfront, agents relied on conversation to extract signals about what clients enjoyed, what made experiences meaningful, and what aspects should be replicated or enhanced in upcoming trips. These narratives provided concrete anchors for identifying underlying constraints and preferences, enabling more personalized recommendations.

\begin{quote}
    \textit{``When I first talked to clients, I always asked them about their past travels. `What was your memorable experience in the past?' ... They told me they had a very unique dining experience in Paris ... [To me] food was then very important [for them].''} (T6)
\end{quote}

\subsubsection{\textbf{A4: Refining Constraints and Preferences}}
This action captures how initially coarse constraints and preferences are transformed into concrete and actionable planning criteria. Rather than being fully specified upfront, many requirements begin as abstract notions. Through exposure to examples, situational evidence, and dialogue, travelers clarify meanings, narrow interpretations, and correct mismatches between intended and interpreted needs.

Travelers often began with only a high-level sense of the constraints they needed to follow, using these as mental reminders rather than concrete specifications. As they encountered more detailed information, these vague notions became finer-grained considerations. For example, when planning travel with her kid, P5 was aware of the need for kid-friendly dining options. As she went through the filtered list of restaurants, she quickly noticed a customer review saying \textit{``the noisiest restaurant [I've] ever been, ''} and she reacted \textit{``OK, so it's like a noisy party. Yep, probably going to scare my child, so not that one.''} Here, an abstract requirement was refined into a concrete exclusion criterion through situational evidence.

From the travel agents' perspective, such moments of vagueness were openings for clarification. Instead of overwhelming clients or risking irrelevant suggestions, agents treated broad or ambiguous terms as openings for dialogue. They sought to uncover the underlying intent by asking follow-up questions and offering detailed variations. As T9 explained, \textit{``If you (client) like to go hiking ... [I will ask is it] easy hiking or paved hikes, or hiking where you're in the dirt, walking through the mountains and stuff like that.''} These clarifying conversations helped bridge the gap between general interests and concrete goals.

AI systems also prompted users to state preferences, but often struggled to support refinement. \gemini prompted users to describe the kinds of sightseeing they wished to visit, which was difficult when preferences were still vague or evolving, while \tripadvisor employed a more guided onboarding process that let users select from predefined categories, as shown in ~\cref{fig:tools}. However, these categories often felt too broad or generic, leading to mismatches between participants' intentions and the AI's interpretations. For example, P1 selected ``nature'' as an interest in \tripadvisor but found none of the recommendations matched his perception of nature. \textit{``Maybe they count beaches as nature. But [that is] not nature for me. So that's when I want to put my specifications,''} said P1. Similarly, P5 added ``short trails'' as an interest but received biking trails as suggestions, noting, \textit{``but this is biking trails. So that's [a] different kind of trails [to what I want].''} In these cases, systems exposed ambiguity but provided limited support for interactively refining meanings.

% However, participants noted that AI-generated options often lacked visible reasoning about how constraints and preferences were applied. As a result, travelers were uncertain about the basis of recommendations and conducted additional verification to assess whether suggested places aligned with their priorities.

\subsubsection{\textbf{A5: Visualizing Constraints and Preferences in the Plans}}
This action externalizes emerging plans into structured representations that make the current plan state inspectable, including how activities are arranged, how fixed or flexible they are, and where constraints and preferences are embedded.

Travelers used structured representations to encode constraint-sensitive information. They recorded times and durations for fixed events, added transportation details, and used colors or icons to highlight important activities (P4, P5, P8). For group trips, participants relied on calendar applications or timetables to record unavailability, such as remote work meetings or children’s nap times. As P4 summarized, these representations helped \textit{``make sure that you are aligning with everyone’s constraints and preferences.''} Together, these practices made time constraints, commitment levels, and coordination requirements visible and easier to revisit as plans developed.

Travel agents similarly encoded constraint status and commitment levels in itineraries. For example, they added highlighted notes indicating reservation confirmations, payment status, or required arrival times, such as \textit{``tour is confirmed and paid for''} (T3), making constraint-sensitive commitments explicit in the plan.

When interacting with AI tools, participants encountered different forms of plan visualization. \gemini typically returned text-based itineraries in response to planning requests. Participants found these difficult to inspect and prompted \gemini to reorganize plans into table-based formats to improve readability. In contrast, \tripadvisor provided an interactive itinerary interface that allowed easier browsing, reordering, and adjustment of activities. These visual formats helped travelers inspect the structure of plans and prepared them for subsequent evaluation.

However, itineraries generated by both AI tools were often presented as uniform step-by-step lists of events, making it difficult for users to distinguish between highlights and secondary activities. Participants noted that this presentation left them uncertain whether their key interests and priorities had been reflected in the plan.

\subsubsection{\textbf{A6: Anchoring High-Priority Constraints and Preferences}}
This action establishes priority structures among constraints and preferences so that key requirements anchor the formation of plans. Rather than treating all considerations equally, anchoring ensures that high-importance and time-sensitive requirements are addressed first, shaping the order in which remaining options are arranged.

From the interviews, we found that travelers coordinated their constraints and preferences (\textit{internal factors}) by prioritizing key events or interests that served as anchors for the plan. For example, P8 listed her most anticipated activities to avoid missing them, building her itinerary by scheduling these first and adding nearby attractions. 
Similarly, P6 anchored his trip around a core goal---visiting an island on the second day---which shaped the rest of his itinerary. 

% Travelers also annotated their own plans to highlight priorities, using techniques such as adding memos or icons next to events (P5), formatting items with different colors (P4), or maintaining a separate list of important tasks (P8). These practices made priorities visible and helped ensure that key interests were not overlooked as plans developed. 

In addition to events strongly reflecting travelers' \textit{internal factors}, decisions that were tied closely to the \textit{external factors}, such as time- and availability-sensitive commitments, were typically addressed first, to secure limited opportunities. This prioritization was further reinforced by common strategies in the travel domain, such as dynamic pricing, where prices fluctuate with demand and availability~\cite{den2015dynamic}.
As T2 explained, even though discussing and finalizing itineraries may take time, they would suggest to first settling these decisions to avoid losing availability or increasing fees by the time the plan is complete.

However, the two AI systems tested in our study lacked such awareness, especially for constraints that were not explicitly specified. For example, when P3 used \gemini to plan a trip around a conference, the system failed to prioritize the conference schedule. Instead, it populated time slots with leisure activities that overlapped with the period when she needed to attend conference sessions. 

% Moreover, itineraries generated by the two AI tools were often presented as step-by-step lists of events, which made it difficult for users to distinguish between highlights and secondary activities. Participants complained that this uniform presentation left them uncertain whether their key interests had been considered. 

\subsubsection{\textbf{A7: Referencing Constraints and Preferences}}
This action makes explicit the rationale behind plan elements by tracing them back to underlying constraints and preferences, especially when plans are generated or proposed by others. It allows travelers to understand why particular activities, accommodations, or schedules were selected, supporting interpretation, trust, and informed revision of emerging plans.

Travel agents routinely performed this action when presenting itineraries to clients. Beyond assembling options, agents highlighted how selected events aligned with expressed interests, prior experiences, and practical constraints. They often adapted default presentation materials to foreground what mattered most to clients. For example, T6 described modifying hotel presentation content to emphasize relevant features:

\begin{quote}
    \textit{``Sometimes it (the default images of the hotel) is like a picture of a meeting room or the dining room, and nobody cares about that. So I'll swap it out for something that I feel is more important. Or if my client says `the spa is really important to me,' then I'll put the spa picture in there.''} (T6)
\end{quote}

This practice made the role of constraints and preferences visible in the composed plan, helping travelers understand why particular options were chosen. On the contrary, travelers complained about the shown pictures of recommended attractions in~\tripadvisor \textit{``were like random pictures about a building or whatever. They were not very informative to me in deciding this''} (P3). Therefore travelers also questioned the relevance of such generated content, especially when no justification was given for why certain places were recommended. 
Rather than simply accepting AI suggestions, they often searched for individual places to verify if they aligned well with their interests through their own research:
\begin{quote} 
\textit{``Because I don't have the information about why they choose these places ... I will try to search all these places to see if I really want to visit or whether they are worth visiting.''} (P6)
\end{quote} 

\subsubsection{\textbf{A8: Evaluating Indivisual Constraints and Preferences}}
This action assesses whether specified constraints and preferences are feasible and realistic on their own. Rather than assuming all stated requirements are valid, evaluation involves identifying misunderstandings, unrealistic expectations, or incomplete requirements that need adjustment before plan composition.

As noted in \cref{external-factors}, travelers sometimes began planning with incomplete knowledge or unrealistic expectations about destinations. Travel agents supported this evaluation by questioning infeasible requests and proposing alternative solutions, helping clients adjust constraints and preferences into workable forms.

In contrast, AI tools rarely questioned unrealistic or underspecified requirements. Instead, they accepted inputs at face value, producing plans that reflected stated preferences even when underlying assumptions were incomplete or impractical. This limited support for evaluating individual constraints often left travelers to discover feasibility issues only after plans were generated.

% \subsubsection{Evaluating Constraints and Preferences}
% This action assesses whether specified constraints and preferences are feasible, consistent, and compatible with each other. Evaluation involves identifying conflicts across options and determining whether constraints and preferences require adjustment before a plan can be composed.

\subsubsection{\textbf{A9: Detecting Conflicts among Constraints and Preferences}}
This action checks consistency across multiple constraints, preferences, and candidate options, identifying contradictions that require resolution. As travelers add or modify requests, interactions among constraints can produce conflicts that are not apparent when considering each constraint in isolation.

As travelers added and modified requests, they sometimes lost track of conflicting details, resulting in contradictory plans that required further resolution. During the study, \gemini often failed to detect such conflicts, leaving travelers with itineraries that overlooked stated constraints. For example, P4 specified that he wanted to stay in and visit the northern part of an unfamiliar destination, but also included a waterfall located in the south as an attraction of interest due to the unfamiliarity of the destination. \gemini generated a plan that included the southern waterfall without reminding P4 of this potential conflict.

% Some implicit conflicts might arise from overfitting the constraints. 
% Passive reliance on user inputs also led to overfitting. 
Conflicts also arose implicitly from overfitting to expressed preferences. In these cases, plans focused too narrowly on a subset of expressed preferences, producing unbalanced itineraries. This frequently occurred with the AI tools in our study, where generated plans emphasized a single type of activity to match stated interests while leaving little room for variety or adjustment, as mentioned in~\cref{integrative-factors}.

\subsubsection{\textbf{A10: Curating Options by Constraints and Preferences}}
This action applies established constraints and preferences to generate, filter, or shortlist candidate plan options. Once constraints and preferences have been surfaced and clarified, they are used as selection criteria to reduce a broad information space into a manageable set of relevant activities, accommodations, or itineraries.

For travelers, curation is often intertwined with ongoing discovery (A2). Rather than separating exploration from selection, participants used emerging preferences as proxies to guide what to search for and what to discard. For example, when planning her trip to Japan, P8 initially browsed travel suggestions from social media. But once she saw an image of red maple leaves, she developed a new interest in autumn foliage, which led her to actively search for attractions where the fall color would still be visible during her travel dates. 
% She described this process as to \textit{``do subtraction''} (P8) to narrow down potential options within the vast landscape of choices.
By browsing websites, reading posts on social media, and consulting other sources, participants unconsciously engaged in information foraging~\cite{pirolli1999information}.
% (i.e., a process of seeking out valuable information while minimizing the time and effort required in the process). 
% travelers gained confidence through collective endorsement (e.g., seeing the same place recommended across sources) or authoritative endorsement (e.g., trusting a travel agent's expertise).

Travel agents curated options after learning clients' needs by assembling tailored sets of recommendations. Rather than presenting exhaustive lists, agents proposed limited alternatives that reflected expressed interests, prior experiences, and practical constraints, allowing clients to focus on evaluating a smaller number of meaningful options.

However, while AI systems also returned a small number of options based on input preferences and constraints, travelers described limitations in how this curation was performed, often due to other poorly executed actions, including not considering that travelers may discover other preferences (A2), or failing to infer implicit preferences (A3) or refine them over time (A4). In \tripadvisor's end-to-end workflow, recommendations were generated directly from initial keywords of interest, effectively fixing the criteria used for selection. As P8 noted, \textit{``because the keywords [of interest] have completely decided what would be included in the plan, what if I figure out something else during planning, do I need to start all over again?''} This made it difficult for travelers to adjust curation as new criteria emerged.

% AI systems also curated options by generating recommendation lists or itineraries from user inputs. However, compared to open-ended online browsing, AI systems provided only a small number of recommendations per request. Participants noted that this constrained exploration, especially in early planning stages when goals and preferences were still forming. As P3 reflected, jumping too quickly into personalized recommendations reduced his awareness of what was available at the destination:

% \begin{quote}
%     \textit{``[Gemini] removed everything else I did not know, ... but on Google [Search], the zoo seems to be like a very famous thing ... I would have never known that there is a famous zoo [in the destination] using AI.''} (P3)
% \end{quote} 

\subsubsection{\textbf{A11: Deferring Decisions through Open Options}}
This action postpones committing to specific plan elements in order to accommodate constraints and preferences that may only become clear during travel. Rather than fully specifying all decisions in advance, deferral preserves flexibility for handling uncertain conditions, emerging opportunities, and situational needs.

Travel unfolded in dynamic real-world contexts shaped by weather, unexpected delays, energy levels, and local availability. To accommodate such contingencies, travelers deliberately left parts of their plans open. For example, P8 avoided deciding on restaurants in advance after experiencing long wait lists on a previous trip. Instead, she planned to explore nearby dining options on-site, allowing her choices to adapt to real-time conditions. 

Similarly, travel agents built in flexibility by preparing alternative recommendations. T6 noted she would prepare options, so that clients \textit{``aren't forced to do every single thing that the whole group is doing.''} This approach personalizes itineraries to individual preferences while accounting for group dynamics, and it ensures that plans remain adaptable for on-the-fly adjustments.

% Travel agents proactively accounted for potential constraints that may arise during the trip. For instance, agents recommended buying refundable tickets when budgets allow, offering flexibility for adjustments or emergencies (T2). Fatigue and jet lag were also key considerations. T3 avoided scheduling fixed reservations immediately after a red-eye flight, allowing clients time to rest. Drawing from years of experience, agents addressed common pitfalls---what T4 described as \textit{``the things that caused us sleepless nights''}---to proactively design smoother itineraries.

% Similar to the interface in \tripadvisor, several travel agents used specific applications to share interactive itineraries with clients, to \textit{``consolidate all the information in one place so they don't have to go to different websites''} (T3). T6 mentioned, \textit{``we have an app ... if something changes and I have to update what time to meet or where to go, I can update the app and the client gets an update on their phone.''}

Our study did not include real-world execution of AI-generated plans, so we could not directly observe how AI systems support contingency management during travel. However, during the interviews, the AI systems sometimes produced fixed itineraries that overfit travelers' stated inputs. This indicates that current AI planning tools primarily support upfront plan generation based on explicit requests, with limited mechanisms for representing uncertainty, deferring decisions, or anticipating situational changes.

\subsubsection{Interplay among Actions}
Although we present these actions as distinct for analytic clarity, they are closely interwoven in practice. For example, visualizing plans (A5) often enables anchoring high-priority events (A6) and conflict detection (A9), while discovering constraints (A2) usually co-occurs with curating options (A10). These patterns reflect how planning work proceeds through iterative adjustments rather than linear steps. 

This interweaving means that breakdowns in one action can propagate to others. When systems fail to identify conflicts or justify decisions based on constraints and preferences, travelers must compensate by performing additional actions themselves, such as independently searching for more options. As our findings suggest, the ability to flexibly navigate interdependencies among actions is central to how humans sustain planning progress—and it is also where current AI tools fall short.
\section{Design Implications: Designing for Evolving Constraints and Preferences in Planning}
We analyzed our empirical findings through the lens of constraints and preferences, and the actions operated on them during planning. Our results showed that in the dynamic process of planning, constraints and preferences are also continuously surfaced, refined, and applied to planning, rather than fixed inputs provided at the outset. This perspective highlighted limitations of current AI-supported planning tools that assume stable and fully articulated user requirements. Based on these findings, we derive design implications for planning tools that support evolving constraints and preferences throughout the planning process. We organize these implications around two aspects: how constraints and preferences should be modeled in an AI planning system, and what functions could be supported to enable users and AI to effectively operate on them during planning.

\subsection{Modeling Evolving Constraints and Preferences}
\label{set:modeling}
We propose that AI-powered travel planning systems should model constraints and preferences as evolving throughout planning. We outline three design considerations for representing and supporting the evolution of constraints and preferences over time.

\subsubsection{Presenting Persistent Constraints and Preferences}
We observed that travelers' constraints and preferences develop alongside their understanding of options and trade-offs. However, most existing planning tools model constraints and preferences as static form inputs or filters, leaving limited room for later adjustment or cohesive management. 

Therefore, planning tools should model constraints and preferences as evolving objects that remain visible and manipulable throughout the planning process. For example, a focus-and-context layout could present recommendations and the plan based on current constraints in the main view, while maintaining visibility of related options in peripheral views. This representation allows travelers to revisit earlier inputs, relax or strengthen constraints, and reorganize priorities as their understanding evolves. 

% Moreover, explicitly representing interactive constraints and preferences helps users develop an understanding of how planning systems reason behind the scenes through exploration and adjustment. When users adjust constraints and immediately see how recommendations or plans change, they gain insights into feasibility boundaries, hidden dependencies, and trade-offs among competing requirements. For example, a travel planning interface could display constraints as adjustable tokens or sliders alongside the plan, where modifying a budget or pacing constraint updates suggested itineraries and visually highlights which parts of the plan are affected or invalid. Such representations make exploration easier with low cost, allowing users to see a broader range of opportunities and plan variations, compare alternatives, and discover possibilities they might not have considered otherwise. 

\subsubsection{Capturing Constraints and Preferences in Context}
The bottom-up discovery of constraints and preferences through information exploration helps users clarify what they want and helps planning tools interpret these needs correctly. However, current systems provide limited means to capture these emerging needs as they arise.

Therefore, travel planning systems should support in-situ capture of user needs, blending information browsing with need discovery. Rather than replacing browsing with summarization, AI should facilitate browsing and interpret interaction cues that arise during exploration. For example, while reading or viewing AI-generated content, travelers might highlight a snippet or image to steer further exploration. AI can further reduce the effort of annotation by detecting potential constraints from user interactions, such as highlighting cost, dates, or activity types, and suggesting them as candidate annotations for travelers to confirm or refine. Such in-situ annotation helps capture emerging constraints without interrupting the flow of exploration and ensures they can be revisited, compared, and used to guide follow-up planning.

% We also found that travelers often express high-level or abstract preferences, such as ``nature'' or ``relaxing trip''. While such abstractions are easy to state, they require refinement to avoid misalignment between users and systems. Current tools provide little support for unpacking abstract constraints into concrete and actionable factors. Planning tools should therefore help users break down abstract constraints into more specific dimensions by leveraging AI's generative and knowledge capabilities. For example, when a traveler searches for ``nature'', the system could present alternative interpretations in-situ such as beaches, national parks, or scenic drives, prompting clarification and shared understanding rather than assuming a single meaning.

\subsubsection{Configuring Interactive Constraints and Preferences}
The dynamic evolution of constraints and preferences requires planning systems to support ongoing configuration and adjustment rather than one-time specification.

Planning systems thus need to represent constraints and preferences as adjustable tokens alongside the evolving plan. Each constraint would carry attributes such as priority and flexibility, allowing travelers to make requirements as strict, negotiable, or exploratory. As travelers adjust a budget or pacing constraint, the system would immediately update itineraries and highlight which parts of the plan become infeasible. This allows travelers to gain insights into feasibility boundaries, hidden dependencies, and trade-offs among competing requirements.

When a traveler expresses an abstract preference such as ``a relaxing nature trip,'' the system could propose alternative interpretations, such as beaches, national parks, or scenic drives. The traveler selects or refines these interpretations, and the system updates constraints and recommendations accordingly. This supports collaborative clarification rather than forcing users or AI to rely on assumed meanings.

\subsection{Supporting Collaborative Actions on Constraints and Preferences}

\begin{table*}[htbp]
  \centering
\caption{Collaborative actions on constraints and preferences. Each action is paired with examples illustrating how users and systems can jointly enact it during planning.}
\label{tab:implication}
  \includegraphics[width=0.98\textwidth]{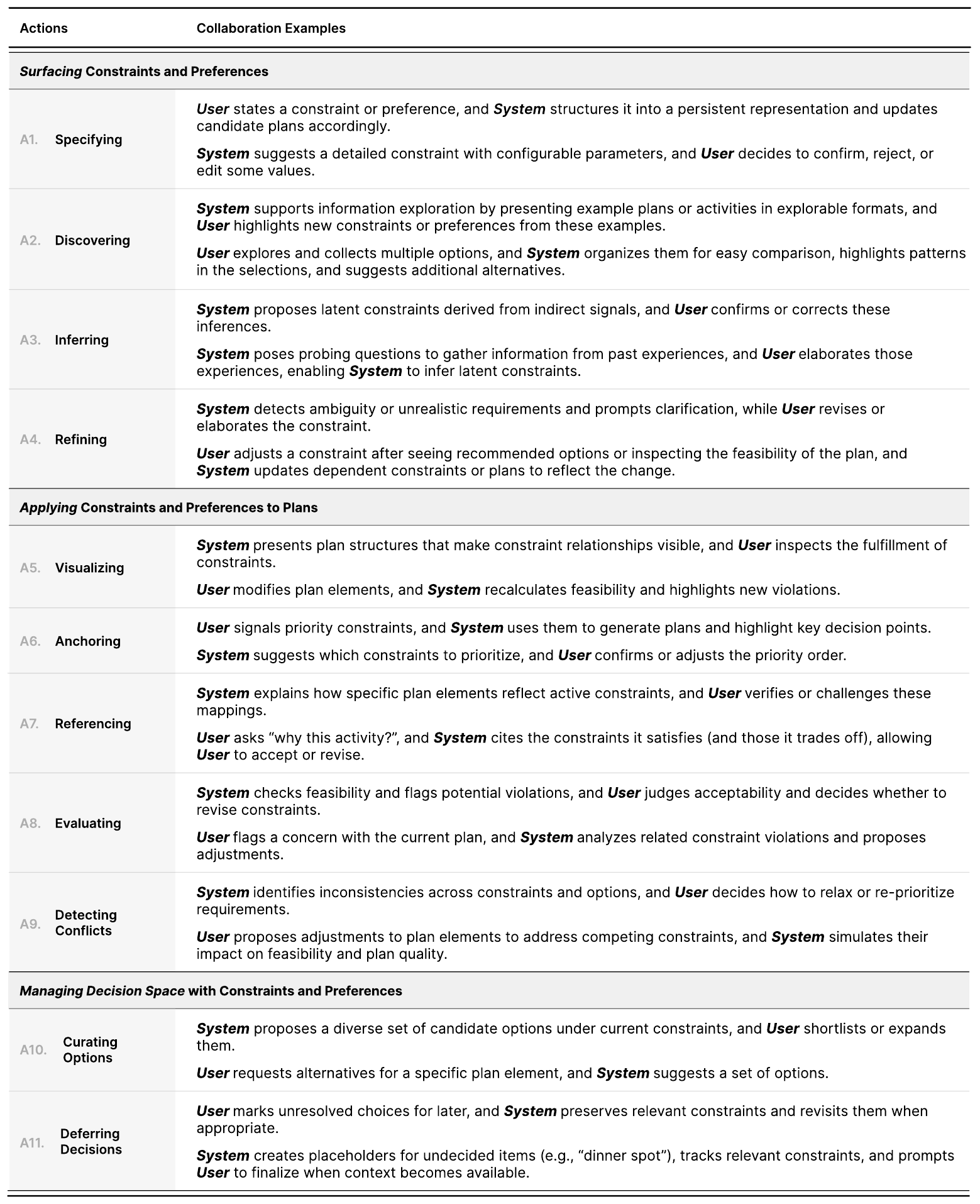}
\Description{Table 3: "Collaborative actions on constraints and preferences," is a two-column table ("Action" and "Collaboration Example") grouping 11 user--system collaborative actions into three planning phases, with each example explicitly labeling the roles "User" and "System": Surfacing constraints and preferences includes Specifying (user states a constraint/preference; system structures it persistently and updates candidate plans, or system proposes a configurable constraint and user confirms/rejects/edits), Discovering (system presents explorable example plans/activities; user discovers constraints, while system organizes collected options for comparison, highlights patterns, and suggests alternatives), Inferring (system proposes latent constraints from indirect signals and shows uncertainty; user confirms/corrects/re-ranks), and Refining (system flags ambiguity or unrealistic requirements and prompts clarification; user revises or adjusts constraints and system updates dependent constraints/plans); Applying constraints and preferences to plans includes Visualizing (system makes constraint relationships visible; user inspects fulfillment and edits plan elements, and system recalculates feasibility and highlights violations), Anchoring (user sets priority constraints; system uses them to guide generation and decision points, or suggests priorities for user to adjust), Referencing (system explains how plan elements map to active constraints; user verifies/challenges, and when asked "why this activity?" system cites satisfied constraints and trade-offs), Evaluating (system checks feasibility and flags violations; user judges acceptability and revises constraints, and system analyzes concerns and proposes adjustments), and Detecting conflicts (system identifies inconsistencies across constraints/options; user relaxes or reprioritizes, proposes changes, and system simulates impacts on feasibility and plan quality); Managing decision space includes Curating options (system proposes diverse candidates under constraints; user shortlists/expands, or requests alternatives for a plan element and system suggests a focused set) and Deferring decisions (user marks unresolved choices for later; system preserves relevant constraints, creates placeholders such as "dinner spot," tracks constraints, and prompts finalization when context becomes available).}
\end{table*}

Our analysis suggests that planning is not only about representing constraints and preferences, but also about acting on them. From our study, we identified a set of recurring actions through which constraints and preferences are identified, applied in planning, and shape the decision space. We examined why and how each action takes place, and compared how travelers, travel agents, and AI systems currently perform them. This comparison showed that several actions are unsupported or only partially handled by current AI tools, limiting users' ability to clarify ambiguous needs, resolve conflicts between constraints, and steer the planning process as preferences evolve.

Therefore, we argue that planning tools should not only model the dynamics of constraints and preferences (\cref{set:modeling}), but also explicitly support collaborative actions on them. This would enable users and AI systems to jointly contribute to evolving requirements throughout planning, reducing the burden on any single party and improving both planning efficiency and outcome quality.

The actions listed in~\cref{set:actions} can be seen as an initial set of functions to support when designing AI travel planning systems. Our findings suggest that these should be implemented as collaborative actions rather than operations performed solely by users or AI. To translate these findings into design considerations, we reformulate the action set as a vocabulary of collaborative actions for human-AI co-planning, which designers can use to structure mixed-initiative interactions around constraints and preferences. 

\cref{tab:implication} illustrates how each action can be jointly enacted by users and AI systems. These collaborative actions are not isolated steps, but often unfold in sequence. For example, evaluating a plan may surface constraint violations that lead to refining requirements, or discovering new preferences during exploration may prompt re-prioritization and regeneration of options. This interconnected structure reflects the iterative nature of real-world planning and highlights the need for systems that support smooth transitions between actions.

\subsection{Summary}
Taken together, these design implications emphasize that effective AI planning support depends not only on representing constraints and preferences, but also on enabling collaborative actions on them. By treating constraints and preferences as shared objects of work and providing explicit interaction support for how they are surfaced, refined, and negotiated, planning tools can better distribute initiative and responsibility between users and AI systems. While grounded in travel planning, these implications offer a transferable lens for designing collaborative AI systems that support evolving requirements in other complex planning domains.

\section{Discussion}
We discuss the generalizability across domains, the implications for understanding human–AI practices, and the limitations of this study. 

\subsection{Beyond Travel Planning}
% Talk about how this can be genrealize to other planning. define the type of planning that are resemble travel planning
% using the lens of constraints ...
While our study focused on travel planning, the dynamics we observed are not unique to this domain. We expect our findings to generalize to planning tasks that are ill-structured~\cite{simon1973structure}, with open-ended goals, evolving constraints, iterative revision, and uncertain environments, rather than tasks with fully specified objectives and fixed constraint models.
% , such as policy design~\cite{dunn2018problem}, urban planning~\cite{AugmentedUrbanPlanning}, and activity scheduling~\cite{RuizActivityScheduling}.
Our findings, therefore, point to a more general perspective on AI-supported planning. Rather than viewing planning systems as tools that transform static inputs into optimized outputs, we argue that planning systems should be designed to participate in the ongoing construction of the planning space. This includes supporting the surfacing of latent considerations, facilitating clarification when goals are underspecified, and helping users reconcile competing constraints as conditions evolve. Across both human-human planning and human-AI planning, progress toward a workable plan depends on collaborative interaction, where different parties introduce, interpret, and refine constraints over time. The collaborative actions we summarized offer a domain-agnostic lens for reasoning about these interactions. Taken together, these insights suggest that designing AI systems to support evolving constraints and collaborative moves is a central requirement for interactive planning systems across domains.

% \subsection{Roles of Constraints and Preferences}
% % constraints usually use to narrow the space, evaluate the results
% % but in this study, we see the role of constraints and preferences as a opening set for exploration. ...

\subsection{Understanding Human-AI Practices}
Recent advances in large language models have accelerated the emergence of AI-powered systems that participate in complex human activities, from travel planning, as examined in this work, to writing, decision making, and collaborative problem solving~\cite {suh2024luminate,park2023choicemates, kim2025plantogether}. For HCI researchers, this raises an important question of how to understand and study effective human–AI practices, beyond evaluating isolated system features or model capabilities~\cite{amershi2019guidelines}.

Our study adopts a comparative framing that examines travelers, professional agents, and current AI systems within the same planning domain. This approach allows us to identify not only what AI systems currently support, but also which actions remain human-led and where collaboration breaks down. By focusing on task-level actions and interaction patterns, we investigate how responsibilities, initiative, and interpretation are negotiated between humans and AI in practice. We hope this methodology may help inform future efforts to study human–AI practices in other AI-for-X systems, where system behavior and human workflows co-evolve over time.

% At the same time, our approach has limitations. We examined how participants used AI through less-structured exploration with exsiting tools and learned from how human-human interactions .. Adopting technology might burst new interaction machanisms taht cannot capture in such ..... 

\subsection{Limitations}

\subsubsection{Participant Sample}
Due to recruitment constraints (yet another constraint), our participant sample was limited to travelers and professional agents within the authors' accessible networks. While we found this to be sufficient to derive meaningful findings, the sample may not fully represent broader populations, such as infrequent travelers or users from different cultural and socioeconomic backgrounds. Future work could expand participant diversity and include cross-cultural or large-scale studies to examine constraints and preferences in varied planning contexts.

\subsubsection{Study Methodology}
We examined how participants used AI through less-structured exploration of existing tools. While this allowed us to capture current practices and expectations, interaction patterns with AI may evolve as users gain familiarity and as systems become more deeply integrated into workflows. Such emerging human–AI practices may not be fully anticipated from present-day human–human collaboration or early-stage AI use. Future work could therefore complement this interview-based approach with longitudinal studies or in-situ deployments to observe how new interaction mechanisms develop over time.

\section{Conclusion}
Constraints and preferences are central to travel planning, yet current AI planning tools often treat them as static inputs rather than evolving and negotiable elements of work. Through interviews with travelers and travel agents, we characterized how constraints and preferences are identified, refined, and applied throughout travel planning. We identified 3 types of constraints and preferences and 11 recurring actions around them during planning and revealed gaps in how current AI systems support travel planning.

Building on these insights, we proposed design implications for AI-powered travel planning systems to model evolving constraints and preferences, and introduced a vocabulary of collaborative actions for structuring interaction between users and AI systems. This treats constraints and preferences as first-class objects for designing mixed-initiative interactions that support discovery, coordination, and revision of requirements over time. While grounded in travel planning, our findings and design implications offer broader lessons for developing AI-powered systems that support complex planning tasks with open-ended goals and evolving requirements.

% We envision AI planning assistants that move beyond generating one-off itineraries to become collaborative partners in dynamic, evolving planning processes. To investigate this vision, we studied how travelers and professional agents navigate shifting constraints and preferences, and how current AI systems succeed and fall short in this space. Our findings highlight the importance of surfacing, clarifying, and reconciling constraints and preferences to build shared understanding and achieve more meaningful plans. From this, we derived design considerations for AI-assisted planning systems that support learning and exploration, enable in-situ elicitation of needs, and flexibly coordinate competing priorities. Building on these insights, we introduced a vocabulary of conversational moves through which users and AI systems work toward establishing shared understanding of external, internal, and integrative constraints and preferences. While not exhaustive, this vocabulary offers a foundation for analyzing how planning unfolds through interaction and for informing the design of systems that enable more flexible and collaborative engagement.

% Together, our empirical findings and design considerations point toward AI planning assistants that are not only more adaptive and responsive, but also more collaborative---helping users construct the planning space, navigate complexity, negotiate trade-offs, and arrive at plans that are both feasible and satisfying.

% % tone down ( oops

\bibliographystyle{ACM-Reference-Format}
\bibliography{__references}

% \appendix

\end{document}